\documentstyle[11pt]{article}
\begin{document}
\begin{center}
{\Large \bf Alpha Clustering and the stellar nucleosynthesis of carbon}\\
[0.7cm]
H.~Oberhummer, H.~Krauss and K.~Gr\"un, T.~Rauscher\\
{\small\it Institut f\"ur Kernphysik, Technische Universit\"at Wien,
A-1040 Vienna, Austria}\\
[0.3cm]
H.~Abele, P.~Mohr and G.~Staudt\\
{\small\it Physikalisches Institut, Univ.~T\"ubingen, D--72076 T\"ubingen,
Germany}\\
\end{center}
{\bf Abstract:} The astrophysical S--factor and reaction rates
  for the triple--alpha
  process are calculated in the direct--capture model. It is shown that
  the stellar carbon production is extremely sensitive to small variations
  in the N--N interaction.


\section{Introduction}
The most important cluster reaction in nuclear astrophysics is certainly
the triple--alpha process occuring in helium burning of stars.
Almost all of the carbon found in our universe
is synthesized by this reaction. According
to most biochemists belief, carbon is the only element
that can form the basis of spontaneously
generated life. Therefore, the stellar production of carbon is
essential for the existence of life in our universe.

In the triple--alpha reaction three alpha particles merge
in a two--step process to form the nucleus $^{12}$C.
In the first step the ground state of $^{8}$Be is formed.
Even so the life time of this nucleus is extremly short
($\approx 10^{-16}$\,s), it is anomalously long compared with the
$\alpha$+$\alpha$ collision time ($\approx 10^{-21}$\,s).
Therefore, the $^{8}$Be is almost in
equilibrium with the $\alpha$--particles. During the life time
of $^{8}$Be a third $\alpha$--particle is captured by
the reaction $^{8}$Be($\alpha$,$\gamma$)$^{12}$C. The
first step depends sensitively on the properties of the
$^{8}$Be ground--state, whereas the reaction rate of the second step
is mainly a consequence of the properties of the  $0_{2}^{+}$ state in
$^{12}$C, which lies
just above the threshold of the $^{8}$Be($\alpha$,$\gamma$)$^{12}$C
reaction.

\section{The triple--alpha process in direct capture}

A detailed account of the calculation and results for the
astrophysical S--factor and the reaction rate of the
triple--alpha process obtained in the direct--capture (DC)
model is given in  \cite{Kra1}.
The $\alpha$--$\alpha$ scattering and the ground--state
properties of $^{8}$Be as well as the bound and resonant states in $^{12}$C,
which are relevant for the triple--alpha process are discussed
in \cite{Sta}.

Direct capture is based on the
description of the dynamics of nuclear processes by a Schr\"odinger
equation with local potentials in the exit channel. The most important
ingredients
of DC are the wave functions for the scattering states in
the entrance channel and the and bound and quasi--bound (resonant)
states in the entrance and
exit channels. The renormalized folding potentials used in the calculation
of the triple--alpha process can be found in  \cite{Sta}.

The astrophysical S--factor for the
reaction $^{8}$Be($\alpha$,$\gamma$)$^{12}$C is calculated
using the DC model \cite{Obe1}, \cite{Moh}, \cite{Kim}
 with the computer code TEDCA \cite{Kra2}.
One observes a resonant contribution of the dominating
$0_{2}^{+}$ state at $0.2875$\,MeV and a nonresonant background.
These two contributions and the interference term are obtained
simultaneously in our model.
Approximating the cross section in the vicinity of the resonance energy
of $E_{\rm R} = 0.2875$\,MeV by a Breit--Wigner formula
we obtain the following widths: (i)
$\Gamma(0_2^+) \approx \Gamma_{\alpha}(0_2^+) = 7.5$\,eV in agreement
with the experimental data $\Gamma(0_2^+)^{\rm exp} = (8.3 \pm 1.0)$\,eV and
(ii)
$\Gamma_{\gamma}(0_2^+) = 4.1$\,meV
which is comparable with the experimental value
$\Gamma_{\gamma}(0_2^+)^{\rm exp} = (3.7 \pm 0.5)$\,meV \cite{Ajz}.

We can now determine the reaction rates for the triple--alpha process
at $10^{8}$\,K using the results for the astrophysical S--factor calculated
using the DC model of the foregoing sections. The
low--temperature rates ($10^{7}$\,K--$10^{8}$\,K)
including the non--resonant parts can be found in \cite{Kra1}.

\section{Variation of the nucleon--nucleon (N--N) interaction}

The following advantages in applying the DC model
together with the folding procedure for the calculation
of the astrophysical S--factor and the reaction rates for the triple--alpha
process are evident.
The first and second step of the process is carried out
consistently using the same model, no parameter
has to be adjusted to the reaction data
and the resonant
and non--resonant part of the astrophysical S--factor
is obtained simultaneously. These features also allow us
to demonstrate the extreme sensitivity of the triple--alpha reaction
to variations of the underlying effective N--N interaction.
A preliminary account of this calculation was
given in \cite{Obe2}.

We now introduce small variations of the strength of the
effective N--N interaction in the calculation of the
$^{8}$Be--ground state and the $0_{2}^{+}$--state in $^{12}$C
without changing the renormalization factors $\lambda$
in the folding potentials given in \cite{Sta}. These factors, introduced
originally to describe higher--order corrections to
the simple folding procedure, are insensitive to
small variations of the effective N--N interaction for
states with a marked cluster structure. This result can be
justified in the single-nucleon exchange approximation \cite{Sat},\cite{Lov}.
In our calculations we also assume that the small
variations of the effective N--N interaction
will only modify the $\alpha$--density distribution needed
in the computation of the folding potentials in higher order.
Furthermore, the changes in the wave function of the final
bound states in $^{12}$C should also be negligible. The
reason for this assumption is that changes of energies and
wave functions are of the order of $|\Delta E/E| \, ^{<}_{\sim} \,
|(\epsilon V_{0})/E|$, where $\Delta E$ is
the change in the energy, $E$ is the energy with respect
to the threshold, $\epsilon$ is the change of the
effective N--N interaction and
$V_{0}$ is the depth of the corresponding folding
potential. Obviously, the above change is much smaller
for states which have a large energy $|E|$
with respect to the threshold.

The changes of the resonance energy of the
ground state of $^{8}$Be as well as the resonance energy
and total width $\Gamma$
of the $0_{2}^{+}$--state in $^{12}$C induced by variations of the
effective N--N interaction are
listed in Table 1. It results that the $\gamma$--width  of the
$0_{2}^{+}$--state
is almost unaffected by this variation. For the calculation
of the reaction rates we will consider in the following
only the resonant contributions. For a temperature
of $1 \cdot 10^{8}$\,K the changes of the reaction rate of the first, second
and both steps in the triple--alpha process are given by $F_{1}$ and $F_{2}$
together with $F_{\rm T} = F_{1} \cdot F_{2}$, respectively and are
listed in the last three columns of Table 1.

As can be seen from this table the reaction
rates of the triple--alpha process are extremely sensitive to small variations
in the
strength of the effective N--N interaction.
A variation of this interaction and
therefore also of the depth of the folding potentials ($V_{0} \approx
100$\,MeV)
by about 0.1\% leads to a shift of the $^{8}$Be ground--state and the
$0_{2}^{+}$--state in $^{12}$C of the order of 10--30\,keV.
These energie shifts introduced in the Boltzmann--factors
$\exp(-E_{\rm R}/kT)$ for the first and second step of the
triple--alpha proces result then in the drastic changes of the
reaction rates shown in Table 1.

{\footnotesize
\begin{table}
Table 1: Effects of
variations in the strength of the effective nucleon--nucleon
interaction on the triple--alpha reaction rate.
$$\vbox{
\halign {\hfil #\enspace\hfil & \hfil #\enspace\hfil & \hfil #\enspace\hfil &
\hfil #\enspace\hfil  &
\hfil #\enspace\hfil & \hfil #\enspace\hfil & \hfil #\enspace\hfil \cr
\noalign{\hrule\smallskip}\cr
& $E^{\rm R}$ [keV]\hfil & $E^{\rm R}$ [keV] \hfil &
$\Gamma$ [eV]\hfil & $F_{1}$\hfil  & $F_{2}$\hfil & $F_{\rm T}$\hfil \cr
& $^{8}$Be$_{\rm g.s.}$\hfil & $0_{2}^{+}$($^{12}$C)\hfil &
 $0_{2}^{+}$($^{12}$C)\hfil & 1.~step\hfil & 2.~step\hfil & total\hfil \cr
\noalign{\smallskip\hrule\smallskip}\cr
\ $0.2\,\%$ & 66.8 & 215.5 & 0.2 & 18 & $3.7 \cdot 10^{3}$ & $6.7 \cdot 10^{4}$
\cr
\ $0.1\,\%$ & 79.2 & 251.7 & 1.5 & 4.2 & 60 & 250 \cr
\ $0\,\%$ & 91.5 & 287.5 & 7.5 &1 & 1 & 1 \cr
\ $-0.1\,\%$ & 103.7 & 323.0 & 29 & 0.24 & 0.016 & 0.0038 \cr
\ $-0.2\,\%$ & 115.7 & 357.9 & 90 & 0.06 & $2.5 \cdot 10^{-4}$ & $1.5 \cdot
10^{-5}$ \cr
\noalign{\smallskip\hrule} \cr }
}$$
\end{table}}

\section{The anthropic principle and the triple--alpha process}

The "anthropic principle" \cite{Car} states that of all
possible universes the one we actually inhabit is tailor--made
for the creation of life. This subject has been extensively
reviewed in \cite{Bar}. The triple--alpha
process plays a key role for the anthropic principle. This can
be seen from the remarkable prediction made by F.~Hoyle
that the creation of carbon would only be
possible, if this reaction would proceed resonantly. This
to our knowledge is the only case, where the anthropic
principle was used to predict the outcome of a
laboratory experiment. Later on,
the prediction of the $0_{2}^{+}$--state in $^{12}$C
was indeed then confirmed
by experiment \cite{Dun}, \cite{Hoy1}, \cite{Hoy2}.

Investigations of the changes in nucleosynthesis of very light
elements by variations of the coupling constants
have already been performed for the unbinding of the
deuteron and binding of the di--proton \cite{Poc}. In this case
the limit was about 16\,\% for the deuteron and about 12\,\% for the
di--proton in order that the anthropic principle is of
significance of such variations.
In the case of the triple--alpha process we showed that
a variation of the coupling constants by only
0.1\,\% would already be of significance for the
anthropic principle, because it reduces the carbon
production by about a factor of 250 at $10^{8}$\,K.

Acknowledgments: We want to thank the Austrian Science Foundation (FWF)
(project P8806-PHY), the \"Osterreichische Nationalbank (project 3924)
and the DFG--project Sta290/2.


\begin{thebibliography}{35}
\bibitem{Kra1} Krauss, H., Gr\"un, K., Herndl, H., Oberhummer, H.,
      Abele, G., Mohr, P., Staudt, G.: submitted to Ap.~J.
\bibitem{Sta} Mohr, P., Abele, H., K\"olle V., Staudt, G.,
      Oberhummer, H., Krauss, H.: this volume
\bibitem{Obe1} Oberhummer, H., Staudt G.: in Nuclei in the Cosmos.
Ed.~H.~Oberhummer, p.~29. Berlin: Springer--Verlag 1991
\bibitem{Moh} Mohr, P., Abele, R. Zwiebel, G., Staudt, G., Krauss, H.,
Oberhummer, H.
Denker A., Hammer J.~W., Wolf, G.:
submitted to  Phys.~Rev.~C
\bibitem{Kim} Kim, K.~H., Park, M.~H., Kim, B.~T.: Phys.~Rev.~{\bf C23}, 363
(1987)
\bibitem{Kra2} Krauss, H.: computer code TEDCA, TU Wien, unpublished
\bibitem{Ajz} Ajzenberg--Selove, F.: Nucl.~Phys.~{\bf A506}, 1 (1990)
\bibitem{Rol} Rolfs, C., Rodney, W.~S.: Cauldrons in the Cosmos.
Chicago: University of Chicago Press 1988
\bibitem{Obe2} Oberhummer, H., Gr\"un, K., Hinterberger, R., Krauss, H.,
Abele, H., Staudt, G., Belyaev, V.: in Proceedings of the Seventh Workshop
in Nuclear Astrophysics 22.--27. March 1993, Ringberg Castle, Bad Wiessee.
Eds.~E.~MŸller, W.~Hillebrandt. Garching: Max-Planck-Institut fŸr Physik und
Astrophysik,
in press
\bibitem{Sat} Satchler, G.~R., Love, W.~G.: Phys.~Rep.~{\bf 55}, 183 (1979)
\bibitem{Lov} Love, W.~G., Owen, L.~W.: Nucl.~Phys.~{\bf A239}, 74 (1975)
\bibitem{Car} Carter, B.: in Confrontation of Cosmological
Theories with Observation, ed.~M.~S.~Longmaier, Dordrecht: Reidel 1974
\bibitem{Bar} Barrow, J.~D., Tipler, F.~J.: The Anthropic Cosmological
Principle. Oxford: Clarendon Press 1986
\bibitem{Dun} Dunbar, D.~N.~F., Pixley, R.~E., Wenzel, W.~A., Whaling, W.:
Phys.~Rev.~{\bf 92}, 649 (1953)
\bibitem{Hoy1} Hoyle, F., Dunbar, D.~N.~F., Wenzel, W.~A., Whaling, W.:
Phys.~Rev.~{\bf 92}, 1095 (1953)
\bibitem{Hoy2} Hoyle, F.: Astroph.~J.~Suppl.~{\bf 1}, 121 (1954)
\bibitem{Poc} Pochet, T., Pearson, J.~M., Beaudet, G., Reeves H.:
Astron.~and Astrophys.~{\bf 243}, 1 (1991)
\bibitem{Liv} Livio, M., Hollowell, D., Weiss, A.~Truran, J.~W.:
Nature {\bf 340}, 281 (1989)
\end{thebibliography}
\end{document}